\documentclass[11pt]{article}

\usepackage[margin=1in]{geometry}
\usepackage{amsmath,amssymb}
\usepackage{hyperref}
\usepackage{enumitem}

\title{The Quantum Rashomon Effect as a Failure of Gluing}
\author{Partha Ghose\thanks{Email: partha.ghose@gmail.com}\\
Tagore Centre for Natural Sciences and Philosophy,\\
Rabindra Tirtha, New Town, Kolkata 700156, India}
\date{}

\begin{document}
\maketitle

\begin{abstract}
Recently Szangolies has argued (in the setting of extended Wigner's-friend scenarios) that
quantum theory permits ``Rashomon'' situations: multiple internally coherent accounts of events that
cannot be combined into a single, consistent global narrative. This note explains why the Rashomon
phenomenon can be understood as a \emph{failure of gluing}: local descriptions over different contexts
exist, but they do not admit a single global ``all-perspectives-at-once'' description. This is the
same mathematical obstruction that underlies modern sheaf-theoretic treatments of contextuality. I then indicate why the same perspective is useful in parts of the
social sciences (quantum-like modelling of cognition, judgment, and decision-making), where ``context
effects'' can likewise be interpreted as the absence of a single joint probability space.
\end{abstract}

\section{The quantum Rashomon effect: a brief summary}
Szangolies \cite{Szangolies2025} has recently strengthened Frauchiger--Renner type arguments
\cite{FrauchigerRenner2018} by showing how different agents can arrive at different, internally
coherent descriptions of ``what happened,'' while no single global narrative can simultaneously
accommodate all of them. The key point is not that any one agent reasons incorrectly. Rather, the
incompatibility arises because the agents' descriptions are tied to \emph{different contexts}\footnote{For physicists: ``context'' may be read as a set of mutually compatible (co-measurable) observables/records; here it is simply the information jointly accessible to an agent.} (different sets of jointly meaningful propositions and records).

For a recent overview of extended Wigner's-friend arguments and how their tensions can be traced to structural assumptions, see Schmid, Y{\i}ng and Leifer \cite{SchmidYingLeifer2023}.

The core feature can be stated without interpretational commitments:
\begin{quote}
There are \emph{locally consistent} descriptions relative to each context, but no \emph{globally consistent}
description that reduces to all of them at once.
\end{quote}
This is precisely the pattern captured by the presheaf/sheaf notion of ``no global section''
\cite{AbramskyBrandenburger2011}. A related viewpoint---measurement as a form of ``sheafification'' of
local, context-indexed truth assignments---is developed in my recent preprint \cite{Ghose}.

\section{Rashomon as failure of gluing (minimal categorical formulation)}
This section uses only the most elementary ``local-to-global'' idea. Readers can keep in mind the
familiar atlas metaphor from geometry: one may describe a sphere using many local maps (charts), but
not every collection of local maps can be glued into a single global map.

\subsection{Contexts and restrictions}
Let $\mathsf{Ctx}$ denote a collection of \emph{contexts}, partially ordered by refinement:
$D \le C$ means that $D$ is a \emph{sub-context} of $C$ (it speaks about fewer jointly meaningful
questions/records, or contains less jointly accessible information). Whenever $D \le C$, there is a
natural \emph{restriction} operation: from any description valid on $C$ one can forget the extra
claims and obtain a description valid on $D$.

\subsection{A presheaf of local accounts}
A convenient way to formalize this is to assign to each context $C$ the set of all admissible
\emph{local accounts} in that context. A convenient bookkeeping device is a \emph{presheaf of local accounts}. For each context $C$,
let $\mathcal{F}(C)$ be the set of coherent accounts one may give \emph{using only the vocabulary/records
available in $C$}. Whenever $D\le C$ (``$D$ is a subcontext of $C$''), there is a restriction map
\[
\rho_{CD}:\mathcal{F}(C)\to \mathcal{F}(D),
\qquad s\mapsto s|_{D},
\]
which simply forgets whatever is not expressible in $D$.
(For readers who like categorical notation: this data is exactly a contravariant functor
$\mathcal{F}:\mathsf{Ctx}^{op}\to \mathbf{Set}$.)

\subsection{Gluing and global sections}
Suppose that a situation of interest is covered by a family of partial perspectives $\{C_i\}$.
A collection of local accounts $s_i\in \mathcal{F}(C_i)$ is called \emph{compatible on overlaps} if,
whenever two perspectives overlap (i.e., share a common subcontext---here $C_i\cap C_j$ denotes this overlap),
they agree on what both can express:
\[
s_i|_{C_i\cap C_j} \;=\; s_j|_{C_i\cap C_j}
\qquad \text{for all } i,j.
\]
A \emph{global section} is a single account $s$ that simultaneously restricts to all the $s_i$:
\[
s|_{C_i}=s_i\qquad \text{for all } i.
\]
When such an $s$ exists, one may legitimately speak of a single global narrative whose restrictions
reproduce all local narratives.

\subsection{The Rashomon pattern}
The Rashomon effect is exactly the failure of this gluing step:
\begin{quote}
There exist locally coherent accounts $s_i\in\mathcal{F}(C_i)$, but there is \emph{no} global section
$s$ whose restrictions reproduce all $s_i$.
\end{quote}
In the sheaf-theoretic approach to contextuality, the \emph{nonexistence of global sections} is the
precise sense in which there is no ``view from nowhere'' assigning outcomes/truth values across all
contexts \cite{AbramskyBrandenburger2011}. (Related local-to-global ideas also appear in topos-based
approaches to quantum theory \cite{DoringIsham2008}.)

There are two conceptually distinct ways ``gluing'' can fail. First, local accounts may already disagree on
overlaps, $s_i|_{C_i\cap C_j}\neq s_j|_{C_i\cap C_j}$; this corresponds to direct influence/disturbance
(context-dependent marginals) and is not the Rashomon phenomenon emphasized here. The stronger, genuinely
contextual obstruction is when the locals are compatible on all overlaps,
$s_i|_{C_i\cap C_j}=s_j|_{C_i\cap C_j}$, yet there is still no global $s\in\mathcal{F}(X)$ extending them.
In the sheaf-theoretic framework this is an \emph{extension obstruction}; after passing to an abelian
coefficient system built from the support of the local data, it can be represented by a nontrivial
\v{C}ech cohomology class (typically in $\check H^{1}$) \cite{Ghose, AbramskyMansfieldBarbosa2012}.

Szangolies cites Lawvere \cite{Lawvere1969} in connection with diagonal/self-reference reasoning.
That is quite natural for Frauchiger--Renner type arguments, which involve agents reasoning about other
agents who also apply quantum theory. 
 
However, Lawvere-style diagonal arguments and gluing obstructions
address different structures:
\begin{itemize}[leftmargin=2em]
\item Diagonal arguments organize fixed-point/self-reference phenomena.
\item Gluing-failure arguments block the passage from local descriptions to a single global description.
\end{itemize}
Thus, the particular categorical tool used by Lawvere has a different target from the one intended here. The Rashomon phenomenon is especially transparent in the local-to-global language of presheaves.

\section{A social-science bridge: quantum-like cognition and decision-making}
The term ``Rashomon'' already belongs to social science and qualitative methodology: multiple accounts
of ``what happened'' can be individually compelling yet mutually incompatible. The quantum case adds a
crisp mathematical template for \emph{when} and \emph{why} a single unified account may not exist.

\subsection{Context effects as ``no single joint distribution''}
In many behavioural and social datasets, responses depend on the measurement context: question order,
framing, priming, and the set of available answers. A classical idealization assumes that there is a
single underlying joint probability space from which all observed probabilities are marginals. When
context effects are strong, no such single joint model may exist (or it may require additional hidden
variables that defeat the intended ``single story''). This is the probabilistic analogue of ``no global
section.''

\subsection{Quantum-like modelling: what it claims (and what it does not)}
Quantum-like models in cognition do \emph{not} require the brain to be quantum-mechanical. The claim is
mathematical: Hilbert-space probability (noncommuting projections, state update, and interference-like
terms) provides a compact way to encode context dependence and order effects in human judgments
\cite{BusemeyerBruza2012,PothosBusemeyer2022}.

A canonical example is the empirically robust \emph{question order effect} in surveys: the probability
of answering ``yes'' to question $A$ can change depending on whether $A$ is asked before or after $B$.
Quantum-like models represent $A$ and $B$ by noncommuting projectors, so that the two orderings are
genuinely different sequential measurements \cite{WangEtAl2014}. In the presheaf language, each
ordering corresponds to a different context, and the failure to embed all contexts into one global
joint model is exactly a gluing obstruction.

\subsection{A useful caution: contextuality vs.\ direct influence}
In behavioural and social-science data, changing the \emph{context} of a question (order, framing,
available information, social setting, \emph{etc.}) can directly change response probabilities.
This is often called \emph{direct influence} (or ``signalling''): even the \emph{marginal} probability of
a response can shift across contexts. Such shifts are ubiquitous and, by themselves, do not yet imply a
deep ``no-go'' obstruction---they may simply reflect that people respond differently under different
conditions.

The \emph{Contextuality-by-Default} (CbD) programme makes this distinction precise by adopting a
conservative modelling rule: a measurement of ``the same property'' in two different contexts is
treated as \emph{two different random variables} \cite{DzhafarovKujala2016CBD,DzhafarovKujala2017}. For example, a response
to question $A$ asked \emph{before} $B$ is represented by a variable $A^{(A\rightarrow B)}$, whereas the
response to $A$ asked \emph{after} $B$ is $A^{(B\rightarrow A)}$. These variables are not assumed equal
\emph{a priori} because their marginals may differ (direct influence).

CbD then defines \emph{noncontextuality} in terms of the existence of a single joint probabilistic model
(\emph{a coupling}) for all context-indexed variables that (i) reproduces the empirically observed joint
distributions within each context, and (ii) aligns ``same-content'' variables across contexts as much as
their different marginals allow. If no such coupling exists---even after allowing for the maximum
possible alignment compatible with the observed marginals---the system is called \emph{contextual} in the
strong, structural sense \cite{DzhafarovKujala2016CBD,DzhafarovKujala2017}.

For the present note, the payoff is conceptual: the slogan ``failure of gluing'' can refer to
different strengths of incompatibility. Ordinary context dependence corresponds to the fact that the
local models (or marginals) change from context to context (direct influence). Strong contextuality is
the sharper statement that \emph{even after accounting for these marginal changes}, there is still no
single global probabilistic description that simultaneously reproduces all the local context-wise data.
This prevents an overly quick identification of every context effect with a foundational obstruction,
and it parallels the physical distinction between mere disturbance/signalling and genuine
Kochen--Specker/Bell-type contextuality.

\section{Concluding remark}
The ``failure-of-gluing'' approach presented above, I hope, will help readers view Szangolies'
Rashomon effect as an instance of contextuality in the strong, local-to-global sense: the obstruction
is structural, not psychological. Conversely, the same formulation offers social scientists a principled
language for modelling context effects without forcing them into the mould of a single joint sample
space.

\section{Acknowledgements}
I acknowledge use of AI tools for editing/grammar verification but all scientific content 
was generated and verified by me.

\end{document}